\documentclass[preprint,endfloats*,superscriptaddress,prb,aps,showpacs,floatfix,citeautoscript]{revtex4}
\usepackage{epsfig,url}
\usepackage{epsf}
\usepackage{graphicx}
\usepackage{textcomp}
\usepackage{amsmath,amssymb}
\usepackage{booktabs}  
\usepackage{dcolumn}
\usepackage{rotating}
\usepackage{bm}        
\usepackage{bbm}       
\newcommand\degrees[1]{\ensuremath{#1^\circ}}

\begin{document}

\title{Identification of fullerene-like CdSe nanoparticles from optical spectroscopy calculations}

\author{Silvana Botti}
\affiliation{Laboratoire des Solides Irradi\'es, CNRS-CEA-\'Ecole Polytechnique,
91128 Palaiseau, France}
\affiliation{European Theoretical Spectroscopy Facility (ETSF)}
\author{Miguel A.\,L. Marques}
\affiliation{Centro de F\'{\i}sica Computacional, Departamento de F\'{\i}sica da Universidade de Coimbra, Rua Larga, 3004-516 Coimbra, Portugal}
\affiliation{European Theoretical Spectroscopy Facility (ETSF)}

\date{\today}

\begin{abstract}
  Semiconducting nanoparticles are the building blocks of optical nanodevices as
  their electronic states, and therefore light absorption and emission, can be
  controlled by modifying their size and shape. CdSe is perhaps the most studied
  of these nanoparticles, due to the efficiency of its synthesis, the high
  quality of the resulting samples, and the fact that the optical gap is in the
  visible range. In this article, we study light absorption of CdSe
  nanostructures with sizes up to 1.5\,nm within density functional theory. We
  study both bulk fragments with wurtzite symmetry and novel fullerene-like
  core-cage structures. The comparison with recent experimental optical spectra
  allows us to confirm the synthesis of these fullerene-like CdSe clusters.
\end{abstract}

\pacs{78.67.Bf, 64.70.Nd, 71.15.Mb}

\maketitle

\section{Introduction}
\label{sec:intro}

For the past years, semiconductor quantum dots have attracted large interest
from the community due to the peculiar role played by quantum
confinement~\cite{bawendi90} and the consequent potential for size-tunable
nanodevices. High quality colloidal nanoparticles can now be routinely
synthesized with a remarkably narrow size distribution (even lower than
5\%~\cite{murray93}), and have already proved to be excellent components for a
variety of applications in opto-electronics (i.e., light-emitting
diods~\cite{coe02}, optically pumped lasers~\cite{tessler02}, photovoltaic
cells~\cite{klimov03}), in telecommunications~\cite{harrison00}, and in
biomedicine as chemical markers~\cite{bruchez98}. Among the colloidal
nanocrystals, CdSe is perhaps the most studied, due to the efficiency of its
synthesis, the high quality of the resulting samples, and the fact that the
optical gap is in the visible range. In most experimental setups, CdSe
nanoparticles are formed by kinetically controlled precipitation, and are
terminated with capping organic ligands, like, e.g., the trioctyl phosphine
oxide (TOPO) molecule, which provide stabilization of the otherwise reactive
dangling orbitals at the surface~\cite{murray93}.

The understanding of the atomic arrangement and surface deformation of CdSe
clusters is quite important. In fact, strong reconstructions imply necessarily
important modifications of the electronic states, and consequently of the
size-dependent optical properties at the basis of all technological
applications.  For ligand-terminated CdSe clusters, both transmission electron
microscopy data~\cite{murray93,shiang95}, molecular dynamics
simulations~\cite{rabani01} and {\it ab initio} structural
relaxation~\cite{puzder04} agree on predicting an atomic arrangement of the
inner Cd and Se atoms analogous to the one in the wurtzite CdSe crystal.  The
extent to which the cluster surface retains the crystal geometry is more
controversial as the surface cannot be easily resolved experimentally.
Generally, if the surface is properly passivated, the reconstruction is assumed
to be small and limited to the outermost layer (and eventually the layer just
beneath it), which is in agreement with molecular dynamics
simulations~\cite{rabani01}. Pudzer {\it et al.}~\cite{puzder04} have predicted
for clusters with diameters up to 1.5\,nm a strong surface reconstruction,
remarkably similar in vacuum and in the presence of passivating ligands.

Recently, the synthesis and probable identification of highly stable
(CdSe)$_{33}$ and (CdSe)$_{34}$ nanoparticles grown in a solution of toluene has
been reported~\cite{kasuya04,kasuya05}. The experimental absorption spectra of
these nanoparticles at low temperature show sharp peaks, similar to the ones
that characterize TOPO-capped clusters of the same size~\cite{murray93}. However,
the surfactant molecules employed in the synthesis process are, in this case,
removed by laser vaporization. Furthermore, an X-ray analysis indicates that the
coordination number of Se is between 3 (the coordination of a fullerene) and 4
(the coordination of the bulk crystal). In view of this, and in absence of
direct structural data, the non-passivated compound nanoparticles are predicted
to have a core-cage structure, composed by a puckered fullerene-like
(CdSe)$_{28}$ cage accommodating a (CdSe)$_n$ ($n$=5,6) wurtzite unit inside.

The core-cage structures proposed by Kasuya {\it et al.}~\cite{kasuya04} are
significantly different from all previous bulk-derived arrangements. This
geometry is proved to be particularly stable by first-principle total energy
calculations~\cite{kasuya04}.  However, up to now no definitive proof has been
given for the identification of the highly stable observed nanoparticles with
the fullerene-like structures.

In this work we address the problem of the identification of CdSe nanoparticles
through the comparison between measured~\cite{kasuya04} and simulated optical
spectra. In fact, as the electronic states are strongly modified by changes of
size and shape, absorption (or emission) peaks are sensitive to such changes as
well. Optical spectroscopy can thus be a powerful tool (especially if it can be
combined with other spectroscopic techniques) to probe the atomic arrangement of
synthesized nanoparticles.

The remaining of this article is organized as follows. In Sect.~\ref{sec:GS} we
discuss the structural relaxation performed within density functional theory
(DFT) for a selection of fragments with wurtzite symmetry and novel
fullerene-like core-cage aggregates.  In Sect. \ref{sec:absorption} we present
calculations of the Kohn-Sham HOMO-LUMO gaps and the absorption spectra for the
optimized geometries, comparing our results with available experimental data.
Finally, in Sect. \ref{sec:concl} we draw our conclusions concerning the
identification of the CdSe core-cage structures.

\section{Ground state calculations}
\label{sec:GS}

\begin{figure}[t]
\begin{center}
\includegraphics[width=8.7cm]{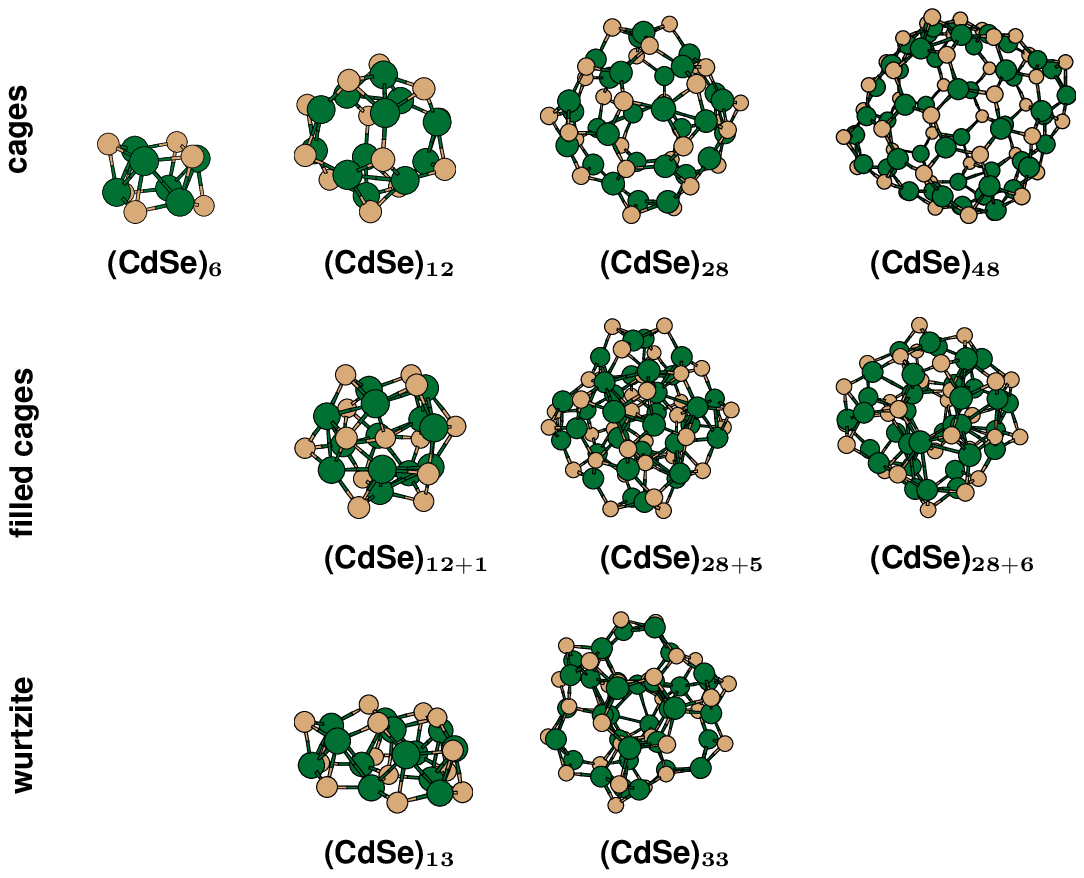}
\end{center}
\caption{\label{fig:structures}
  (Color online) Relaxed cages, filled cages and wurtzite structures of
  (CdSe)$_n$. Cd atoms are in green and Se atoms are in beige.}
\end{figure}

The atomic positions of our CdSe nanoparticles were obtained by geometry
optimization within DFT~\cite{siesta}, using norm-conserving
pseudopotentials~\cite{troullier91} and the local density approximation (LDA) for the
exchange-correlation potential~\cite{LDA}. For all calculations presented here we
used 2 Cd (5$s^2$) and 6 Se (4$s^2$4$p^4$) valence electrons. We verified that
using a pseudopotential with 18 valence electrons for Cd did not change
significantly the relaxed bond lengths and angles.

We considered initial candidate structures with different sizes ranging up to
about 1.5\,nm. To build up these atomic arrangements we started from three
different kinds of ideal geometries: (i) bulk fragments cut into the infinite
wurtzite crystal, (ii) octahedral fullerene-like cages made of four and
six-membered rings and (iii) the core-cage structures of
Ref.~[\onlinecite{kasuya04}], composed of puckered CdSe fullerene-type cages which
include (CdSe)$_n$ wurtzite units of adequate size to form a three-dimensional
network.  We assume that the Cd-Se distance before structural relaxation is the
distance in the CdSe wurtzite crystal, calculated within DFT in the same
approximations used for the nanoparticles~\cite{siesta}: its value (0.257\,nm)
compares well with the experimental value (0.263\,nm).

\begin{figure}[t]
\begin{center}
\includegraphics[width=8.5cm]{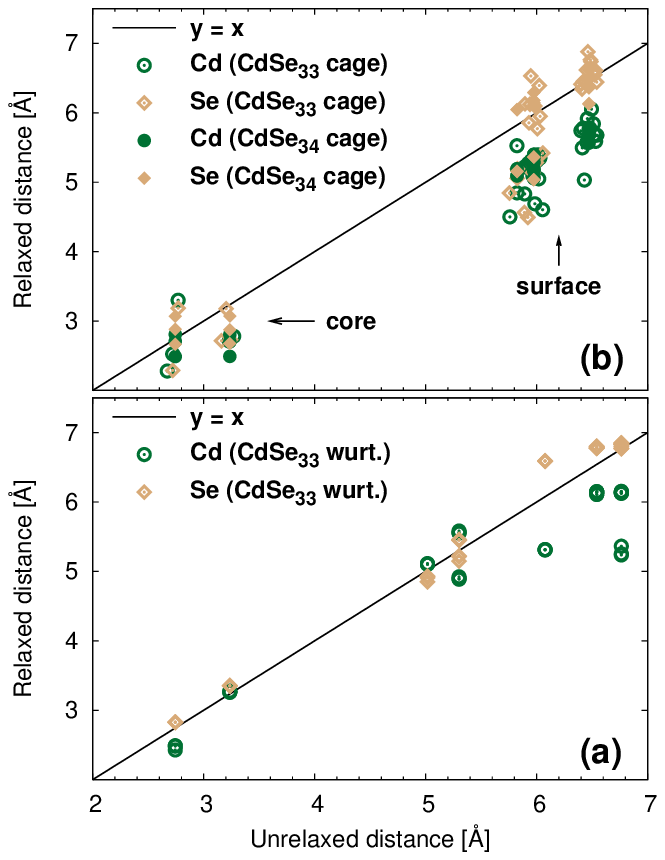}
\end{center}
\caption{\label{fig:distances}
  (Color online) Distance of Cd atoms (circles) and Se atoms (diamonds)
  from the center of the cluster after geometry optimization, as a function of
  their distance before optimization. An atom that lies on the straight line
  $y=x$ has not changed its position. In panel (a) results of the analysis for
  (CdSe)$_{33,34}$ core-cage clusters, in panel (b) for the (CdSe)$_{33}$ wurtzite
  cluster.  }
\end{figure}

\begin{figure}
  \begin{center}
    \includegraphics[width=7.5cm]{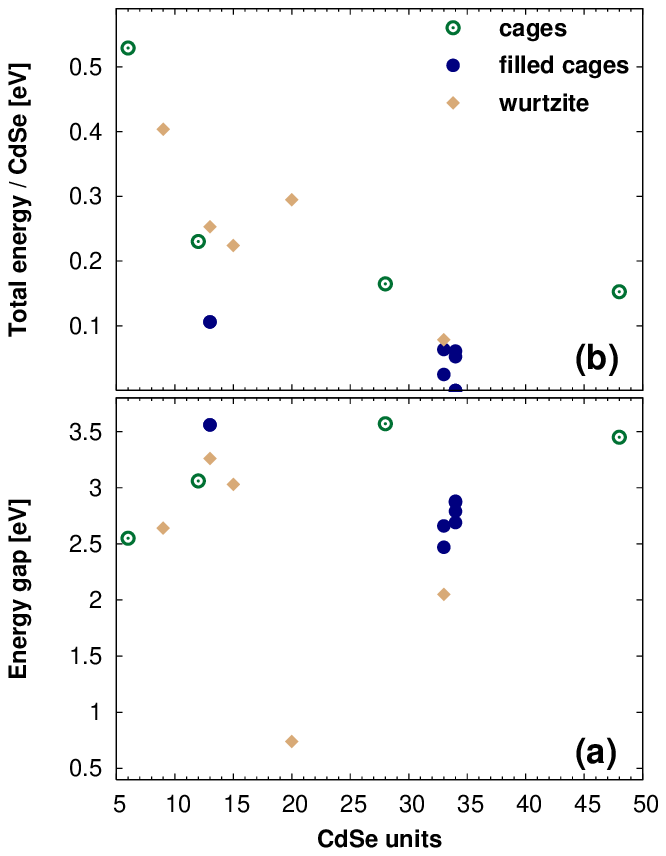}
  \end{center}
\caption{\label{fig:etotal-gap}
  (Color online) Calculated total energies per CdSe unit (a) and HOMO-LUMO gaps
  (b) as a function of the number of CdSe units. The zero of energy is set to
  the total energy per pair of the (CdSe)$_{34}$ core-cage. The empty (filled)
  circles refer to cage (core-cage) clusters, while the diamonds refer to
  wurtzite-based structures. }
\end{figure}

Atomic arrangements after optimization are depicted in
Fig.~\ref{fig:structures}. All clusters suffer contraction upon geometry
minimization. For example, (CdSe)$_{33,34}$ clusters experience a size reduction
of about 1--1.5\%, in agreement with the X-ray analysis of
Ref.~[\onlinecite{kasuya04}]. However, as the relaxation affects mainly the
outermost atoms, the overall effect is more pronounced in smaller structures,
where the average Cd-Se distance decreases up to 4\%. This contraction does not
conserve the overall shape, as Cd atoms are pulled inside the cluster and Se
atoms are puckered out. As a consequence, Cd-Cd average distances can be reduced
by 30\%, while Se-Se distances remain essentially unvaried. This is clearly
visible in Fig.~\ref{fig:distances}, where the relaxed distance of Cd (circles)
and Se (diamonds) atoms from the center of the cluster is plotted for
(CdSe)$_{33,34}$ clusters as a function of their distance before relaxation. If
the atoms remained in their initial position, all datapoints would fall on the
straight line $y=x$.  The fact that most Cd atoms lie below the line, while
most Se atoms are above it, shows that in our simulation Cd atoms prefer to
move inward and Se atoms outward. That puckering happens independently of the
cluster size and it is in agreement with previous
calculations~\cite{kasuya04,puzder04}.

All wurtzite fragments get significantly distorted upon relaxation and break
their original symmetry. In agreement with the findings of Pudzer {\it et
  al.}~\cite{puzder04}, the strong modification of bond lengths and angles
concerns essentially the surface layer. In particular, we can see in
Fig.~\ref{fig:distances}(a) that the wurtzite-type (CdSe)$_{33}$ is already
large enough to conserve a bulk-like crystalline core. In fact, the spread of
the points from the straight line is pronounced only for the external shell of
atoms. The calculated overall contraction of the cluster is consistent with
experimental data~\cite{zhang02}.  Also the empty cages [(CdSe)$_{12}$,
(CdSe)$_{28}$, and (CdSe)$_{48}$] get puckered, but conserve their overall
shape. Their binding energies are smaller by 0.1--0.2\,eV with respect to the
binding energies of the corresponding filled cages [see Fig.\ref{fig:etotal-gap}(b)], 
showing the importance of preserving the three-dimensional $sp^3$ Cd-Se network. To optimize the core-cage
structures [(CdSe)$_{12+1=13}$,(CdSe)$_{28+5=33}$, and(CdSe)$_{28+6=34}$] we
created different starting arrangements assuming different orientations for the
encapsulated CdSe$_{n=1,5,6}$ units. In the relaxed assemblies the distributions
of bond lengths and angles result very similar despite of the distinct initial
configurations. Nevertheless, we decided to use few distinguishable relaxed
geometries in the spectroscopy calculations in order to evaluate the dependence
of the optical spectra on small changes in the atomic positions.

In Fig.~\ref{fig:etotal-gap}(a) we summarize our results for the total energy
per CdSe unit in the configurations studied here. The zero of energy is set to
the lowest energy structure of (CdSe)$_{34}$.  The core-cage (CdSe)$_{33}$
cluster is significantly more deformed under optimization than (CdSe)$_{34}$,
but it turns out to have a very similar binding energy. We thus confirm that the
fullerene-like geometries are particularly stable~\cite{kasuya04}.  The filled cage
structure made of 13 units gives as well a relative minimum in the total energy
per pair.  In the case of (CdSe)$_{13}$ and (CdSe)$_{33}$ it is possible to
compare the total energies of the different three-dimensional isomers: the
core-cage nanoparticles have a slightly higher binding energy [0.15\,eV for
(CdSe)$_{13}$ and 0.05\,eV for (CdSe)$_{33}$].  However, we should not forget
that the energy differences we are discussing here are all very tiny, sometimes
of the same order of magnitude as the accuracy of the calculations. That fact
confirms how difficult it can be to extract structural information from a single
number (the total energy) and leads to the conclusion that the simple analysis
of total energy differences cannot be conclusive to demonstrate the existence of
fullerene-like CdSe clusters.

\section{Optical absorption}
\label{sec:absorption}

\begin{figure}
  \begin{center}
    \includegraphics[width=7.5cm]{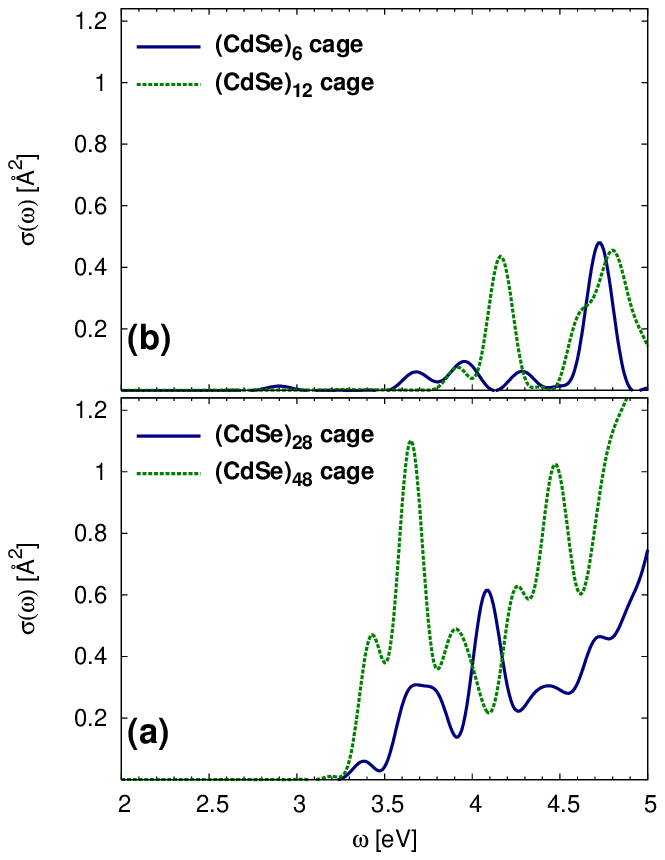}
  \end{center}
  \caption{\label{fig:abs-cages}
    (Color on line) Photoabsorption cross section $\sigma(\omega)$ of the empty cages. In panel
    (a) the absorption spectra of (CdSe)$_{48}$ (dotted line) and (CdSe)$_{28}$
    (solid line); in panel (b) the absorption spectra of (CdSe)$_{12}$ (dotted
    line) and (CdSe)$_{6}$ (solid line). }
\end{figure}

\begin{figure}
  \begin{center}
    \includegraphics[width=7.5cm]{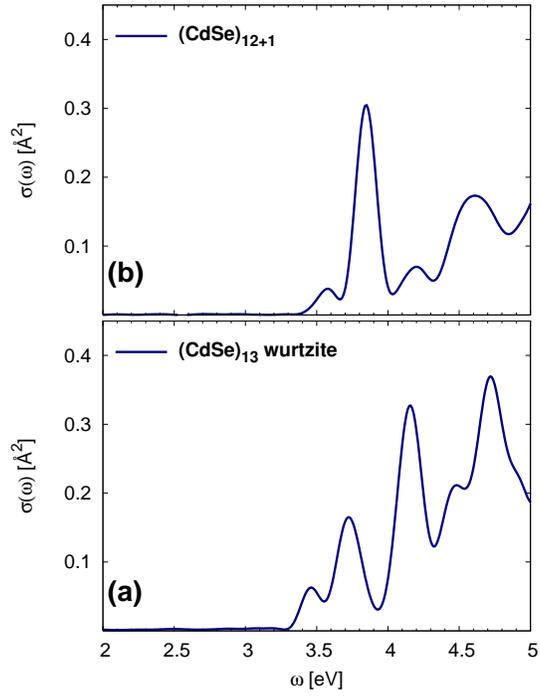}
  \end{center}
  \caption{\label{fig:abs-13}
    (Color online) Photoabsorption cross section $\sigma(\omega)$ of the isomers of
    (CdSe)$_{13}$.  In panel (a) the absorption spectrum of the wurtzite
    structure; in panel (b) the absorption spectrum of the filled cage.}
\end{figure}

\begin{figure}[t]
  \begin{center}
    \includegraphics[width=7.5cm]{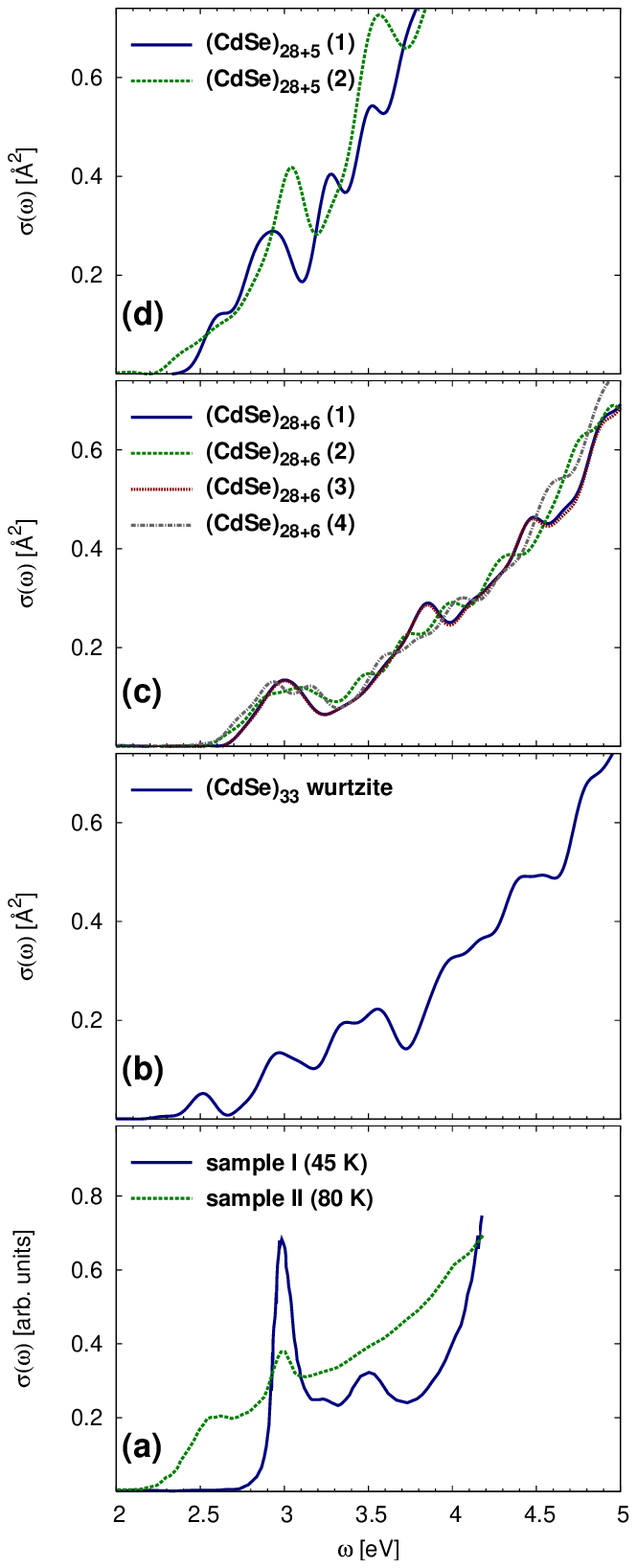}
  \end{center}
  \caption{\label{fig:abs-33} 
    (Color on line) Photoabsorption cross section $\sigma(\omega)$ of the isomers of
    (CdSe)$_{33,34}$. The experimental data in arbitrary units (a) are compared
    with our calculated spectra (b), (c), and (d). The different curves in
    panels (c) and (d) correspond to distinct relaxed geometries obtained
    starting from different filled cages.}
\end{figure}

From the relaxed geometries we obtain the optical spectra at zero temperature
using time-dependent density functional theory (TDDFT)\cite{runge84} in the
adiabatic local density approximation~\cite{gross85}. Of course, calculations
can be predictive only if the theoretical approach employed guarantees a precise
assignment of the peak positions. {\it Ab initio} spectroscopy calculations in
TDDFT have already proved to be a reliable and efficient tool to characterize
structural transitions in C and B clusters~\cite{castro02,marques05}. In the
same implementation~\cite{octopus} used here, TDDFT has been applied to study
metal and semiconducting clusters~\cite{castro04,marques05}, aromatic
molecules~\cite{yabana99}, protein chromophores~\cite{marques03}, etc.,
reproducing the low energy peaks of the optical spectra with an average accuracy
below 0.2\,eV~\cite{castro04}. The accuracy in reproducing transitions of
intermediate energy is known to be somewhat deteriorated, due to the wrong
asymptotic behavior of the LDA exchange-correlation potential. For this reason,
we will focus the analysis of our results on the lowest energy peaks.

In Fig.~\ref{fig:abs-cages} we display the photoabsorption spectra of the empty
cages of different diameters. First of all, we observe that the absorption
threshold is systematically blueshifted with respect to the bulk optical gap
($\simeq$ 1.8 eV).  In particular, due to quantum confinement effects, the shift
is more pronounced when the cluster size is smaller.  Second, we can compare the
absorption threshold with the Kohn-Sham gap between the highest occupied and
lowest unoccupied molecular orbitals (HOMO-LUMO) reported in
Fig.~\ref{fig:etotal-gap}(a): the Kohn-Sham gap is systematically smaller than
the TDDFT absorption threshold.  It is well known that the simpler approach of
taking the differences of eigenvalues between HOMO and LUMO orbitals gives peaks
at lower frequencies in complete disagreement with the experimental
spectra~\cite{castro02}.  We note that the TDDFT optical gaps include both
electron-electron and electron-hole corrections to the Kohn-Sham gap at the
level of the adiabatic local density approximation.

We should keep in mind that the opening of the gap due to confinement can be
counterbalanced by a closing of the gap due to surface reconstruction.  This
leads to a non trivial dependence of the absorption gap as a function of the
cluster size. This effect is already present at the Kohn-Sham level [see
Fig.\ref{fig:etotal-gap}(a)] and it persists in TDDFT spectra.  In fact, the
calculated absorption curves are strongly dependent not only on the cluster size
but also on the details of its atomic arrangement.  This is evident if we
compare the optical response of the different isomers of (CdSe)$_{13}$ in
Fig.~\ref{fig:abs-13} and of (CdSe)$_{33}$ in Fig.~\ref{fig:abs-33}. The
absorption threshold is lower in wurtzite-type clusters as the HOMO-LUMO gap is
reduced due to the presence of defect states in the gap as a consequence of the
strong surface deformation. For a similar reason, the larger surface deformation
of the core-cage (CdSe)$_{33}$ aggregate in comparison with the more stable
(CdSe)$_{34}$ structure explains why the first starts absorbing at lower
energies than the second.  Finally, we note that the similar curves of different
colors in panels (c) and (d) of Fig.~\ref{fig:abs-33} correspond to distinct
core-cage geometries obtained in various optimization simulations. We conclude
that the dependence of the relevant peak positions and shapes on the different
atomic arrangements is not negligible, but the peak positions and oscillator
strengths are sufficiently defined for our purposes.

A comparison with measured spectra~\cite{kasuya04} is possible for nanoparticles
made of $33$ and $34$ CdSe units (see Fig.~\ref{fig:abs-33}).  The solid blue
line in panel (a) of Fig.~\ref{fig:abs-33} refers to room temperature absorption data for mass-selected nanoparticles
prepared in toluene at \degrees{45}C (sample I), while the green dotted line
correspond to analogous data for the solution prepared at \degrees{80}C 
(sample II). Both samples are characterized by strong absorption at 3\,eV. For
sample II the experimental data show the appearance of a broad peak extending to
lower energies. This peak turns out to move to even lower energies when the
temperature and the time in the synthesis process increase.  In a simple quantum
confinement picture, these findings suggest that larger particles, possibly
reconstructed bulk fragments, are formed when the temperature increases.
Moreover, the sharp peak at about 3\,eV, which is always present, was
hypothesized to be the signature of the highly resistant fullerene-like
clusters.

Our calculated spectra shown in Fig.~\ref{fig:abs-33} prove the presence of fullerene-like core-cage
structures. The theoretical optical response of all our model core-cage
(CdSe)$_{34}$ clusters is indeed characterized by a well defined absorption peak
at 3\,eV. Also the core-cage (CdSe)$_{33}$ cluster and the (CdSe)$_{33}$
reconstructed bulk fragment can contribute to this peak. However, they cannot be
present in sample I, as that would be signalled by the appearance of a broader
peak at lower energy, which is absent in the experimental spectrum. On the other
hand, a peak at about 2.5\,eV, connected to the peak at 3\,eV by a region of
increasing absorption, is present in the spectrum for sample II. Our
calculations show that the (CdSe)$_{33}$ wurtzite fragment is responsible for
the peak at 2.5\,eV, while the broad absorption region between 2.5\,eV and 3\,eV
can be explained by the presence of (CdSe)$_{33}$ core-cage structures. This is
in disagreement with the intuition of Ref. \cite{kasuya04} that bulk
fragments of about 2.0\,nm gave rise to the broad absorption below 3\,eV.
Although it is true that a peak in that energy range is typical of 2.0\,nm
crystalline-like TOPO-passivated nanoparticles~\cite{murray93}, the effect of
the structural deformation of ligand-free samples is more than a simple
broadening of the absorption line.  In fact, ligand-free cluster can absorb at
lower energies than passivated cluster of the same size due to a closing of the
gap induced by surface reconstruction. On the other hand, the experimental
evidence of a continuous redshift of the 2.5\,eV peak when the temperature and
the time of the synthesis process increase can only be explained by the gradual
addition of atoms to the (CdSe)$_{33}$ bulk fragment, as the creation of the
next stable core-cage structure would require filling the successive
fullerene-like cage made of 48 CdSe units.

\section{Conclusions}
\label{sec:concl}

In summary, we performed an ab-initio study of CdSe nanoparticles, considering
different cluster sizes and atomic arrangements. Upon geometry optimization,
these clusters get slightly contracted and puckered, with Cd atoms pulled in and
Se atoms pushed out. The most stable structures were, as expected from previous
results, fullerene-like clusters including a wurtzite core. Optical spectra
calculated with TDDFT turn out to be blue-shifted with respect to the bulk gap,
due to confinement effects, and red-shifted with respect to the Kohn-Sham
HOMO-LUMO gaps, due to the inclusion of electron-electron and electron-hole interactions
at the level of the adiabatic LDA. As the opening of the gap due to confinement can be
counterbalanced by a closing of the gap due to surface reconstruction, we found
a non trivial dependence of the absorption gap as a function of the cluster size.
Furthermore, by comparing our theoretical spectra with measurements, we could prove the
existence of the stable core-cage fullerene-like structures recently
hypothesized.

\section{Acknowledgments}

We would like to thank A. Rubio for many useful discussion. We further
acknowledge partial support by the EC Network of Excellence NANOQUANTA
(NMP4-CT-2004-500198). Computations were performed at the Institut du
D\'eveloppement et des Ressources en Informatique Scientifique (IDRIS, France,
project \#020544) and at the Laborat\'orio de Computa\c{c}\~ao Avan\c{c}ada of
the University of Coimbra (Portugal).


\end{document}